\documentclass[prb,twocolumn,showpacs,preprintnumbers,amsmath,amssymb]{revtex4}

\usepackage{graphicx}
\usepackage{dcolumn}
\usepackage{bm}

\preprint{submitted to the Physical Review B}

\begin{document}

\title{Exchange interactions and temperature dependence of
the magnetization  in half-metallic Heusler alloys}

\author{E.~\c Sa\c s\i o\~glu}\email{ersoy@mpi-halle.de}
\affiliation{Max-Planck-Institut f\"ur Mikrostrukturphysik,
D-06120 Halle, Germany}

\author{L. M. Sandratskii}\email{lsandr@mpi-halle.de}
\affiliation{Max-Planck-Institut f\"ur Mikrostrukturphysik,
D-06120 Halle, Germany}

\author{P. Bruno}\email{bruno@mpi-halle.de}
\affiliation{Max-Planck-Institut f\"ur Mikrostrukturphysik,
D-06120 Halle, Germany}

\author{I. Galanakis}\email{i.galanakis@fz-juelich.de}
\affiliation{Institut of Microelectronics, NCSR ``Demokritos'',
15310 Aghia Paraskevi, Athens, Greece}

\date{\today}

\begin{abstract}
We study the exchange interactions in half-metallic Heusler alloys
using first-principles calculations in conjunction with the
frozen-magnon approximation. The Curie temperature is estimated
within both mean-field (MF) and random-phase-approximation (RPA)
approaches. For the half-Heusler alloys NiMnSb and CoMnSb the
dominant interaction is between the nearest Mn atoms. In this case
the MF and RPA estimations differ strongly. The RPA approach
provides better agreement with experiment. The exchange
interactions are more complex in the case of full-Heusler alloys
Co$_2$MnSi and Co$_2$CrAl where the dominant effects are the
inter-sublattice interactions between the Mn(Cr) and Co atoms and
between Co atoms at different sublattices. For these compounds we
find that both MF and RPA give very close values of the Curie
temperature slightly underestimating experimental quantities. We
study the influence of the lattice compression on the magnetic
properties. The temperature dependence of the magnetization is
calculated using the RPA method within both quantum mechanical and
classical approaches.

\end{abstract}


\maketitle

\section{Introduction}\label{sec1}

During the last decade the half-metallic ferromagnets have become
one of the most studied classes of materials. The existence of a
gap in the minority-spin band structure leads to 100\%
spin-polarization of the electron states at the Fermi level and
makes these systems attractive for applications in the emerging
field of spintronics. \cite{Zutic} In half-metals the creation of
a fully spin-polarized current should be possible that should
maximize the efficiency of magnetoelectronics
devices.\cite{deBoeck}

The half-metallicity was first predicted by de Groot and
collaborators in 1983 when studying the band structure of a
half-Heusler alloy NiMnSb.\cite{deGroot} They found that the
spin-down channel is semiconducting. In 2002 Galanakis \textit{et
al.} have shown that the gap arises from the interaction between
the $d$-orbitals of Ni and Mn creating bonding and antibonding
states separated by a gap. \cite{GalanakisHalf} Ishida and
collaborators  have proposed that also the full-Heusler compounds
of the type Co$_2$MnZ, where Z stands for Si and Ge, are
half-metals.\cite{Ishida} In these compounds the origin of
half-metallicity is more complex than in the half-Heusler alloys
because of the presence of the states located entirely at the Co
sites.\cite{GalanakisFull} Several other Heusler alloys have been
predicted to be half-metals.\cite{various}  Akinaga and
collaborators \cite{Akinaga} were able to crystallize a CrAs thin
film in the zinc-blende structure, that is similar to the lattice
of the Heusler alloys. The magnetic moment per formula unit was
found to be close to 3$\mu_B$ that corresponds to the integer
value characteristic for half-metals. A number of further
half-metallic materials are CrO$_2$ in a metastable cubic phase,
Fe$_3$O$_4$, the manganites (\textit{e.g}
La$_{0.7}$Sr$_{0.3}$MnO$_3$)\cite{Soulen}, the diluted magnetic
semiconductors (\textit{e.g.} Mn impurities in Si or GaAs).
\cite{Freeman,Akai}

Besides strong spin polarization of the charge carriers in the
ground state the spintronics materials must possess a high Curie
temperature to allow the applications in the devices operating at
room temperature. Available experimental information shows that
the Heusler alloys are promising systems also in this respect.
\cite{Webster} Up to now the main body of the theoretical studies
was devoted to the properties  of the half-metallic
gap.\cite{Nanda} Recently, Chioncel and collaborators studied the
influence of the correlation effects on the electron structure of
CrAs.\cite{Chioncel} They found that the spin-magnon interaction
leads to the appearance of non-quasiparticle states in the
spin-minority channel. The states are shown to lie above the Fermi
level and to be sensitive to the value of the lattice constant.
For a number of Heusler alloys it was shown that half-metallicity
is preserved under tetragonalization of the crystal lattice
\cite{Block} and application of the hydrostatic pressure
\cite{Picozzi-Gala}. Mavropoulos \textit{et al.} studied the
influence of the spin-orbit coupling on the spin-polarization at
the Fermi level and found the effect to be very small
\cite{Mavropoulos} that is in agreement with a small orbital
moment calculated by Galanakis. \cite{GalanakisOrbit} Larson
\textit{et al.}\cite{Larson} have shown that the structure of
Heusler alloys is stable with respect to the interchange of atoms
and Orgassa and collaborators and Picozzi and collaborators have
demonstrated that a small degree of disorder does not destroy the
half-metallic gap.\cite{Orgassa,Picozzi2004} Dowben and Skomski
have shown that at non-zero temperatures the spin-wave excitations
lead to the presence at the Fermi level of the electron states
with opposite spin projections leading to decreasing
spin-polarization of the charge carriers. \cite{Dowben}

Despite very strong interest to the half-metallic ferromagnetism
in Heusler alloys the  number of theoretical studies of exchange
interactions and Curie temperature  in Heusler alloys is still
very  small. The first contribution to the density functional
theory of the exchange interactions in these systems was made in
an early paper by K\"ubler {\textit et al.} \cite{Kubler83} where
the microscopic mechanisms of the magnetism of Heusler alloys were
discussed on the basis of the comparison of the ferromagnetic and
antiferromagnetic configurations of the  Mn moments. Recently, the
studies of the inter-atomic exchange interactions in several
Heusler compounds were reported by the present authors and
Kurtulus \textit{et al.}
\cite{Sasioglou2004,Sasioglou2005,Kurtulus}. ~\c Sa\c s\i o\~glu
\textit{et al.} studied the exchange interactions in
non-half-metallic Ni$_2$MnZ (Z=Ga,In,Sn,Sb) and half-metallic
Mn$_2$VZ (Z=Al,Ge). The importance of the inter-sublattice
exchange interaction has been demonstrated. For example, in the
case of Mn$_2$VZ (Z=Al,Ge) it was shown that the antiferromagnetic
coupling between the V and Mn moments stabilizes the ferromagnetic
alignment of the Mn moments. K\"ubler\cite{Kubler2003} estimated
$T_C$ of NiMnSb to be 601 K to 701 K depending on the approach
used in the calculations. These values are in good correlation
with experimental value of 730 K.\cite{Webster}

The main task of the present contribution is the study of the
exchange interactions in both half- and full-Heusler alloys. We
use the calculated exchange parameters to estimate the Curie
temperature in both the random phase ($T_C^{RPA}$) and the mean
field approximations ($T_C^{MFA}$). In Section \ref{secII} we
briefly discuss the formalism employed in the calculations. In
Section \ref{secIII} we present the results on the spin magnetic
moments and the density of states (DOS) for four compounds
studied: NiMnSb, CoMnSb, Co$_2$MnSi and Co$_2$CrAl. In Section
\ref{secIV} we discuss the calculated exchange interactions and
Curie temperatures. Section \ref{secV} is devoted to the
consideration of the temperature dependence of magnetization. The
films of Heusler alloys grown on different substrates can have
different lattice parameters and, as a result, noticeable
variation of the electron structure. Section \ref{secV} contains
the summary. In the Appendix, we present the formalism for the
calculation of the Curie temperature of a multi-sublattice
ferromagnet within the framework of the random phase
approximation.

\begin{figure}
\includegraphics[scale=0.3]{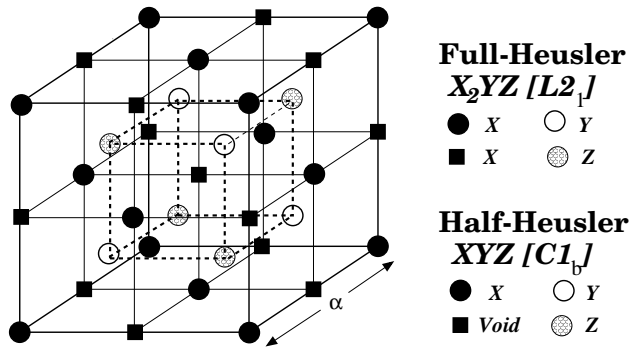}
\caption{$C1_b$ and $L2_1$ structures adapted by the half- and
full-Heusler alloys. The lattice consists from 4 interpenetrating
fcc lattices. In the case of the half-Heusler alloys (XYZ) one of
the four sublattices is vacant. If all atoms were identical, the
crystal structure would be a simple bcc lattice} \label{fig1}
\end{figure}

\section{Calculational Method}\label{secII}

Half- and full-Heusler alloys crystallize in the $C1_b$ and $L2_1$
structures respectively (see Fig. \ref{fig1}). The lattice
consists from 4 interpenetrating fcc lattices. In the case of the
half-Heusler alloys (XYZ) one of the four sublattices is vacant.
The Bravais lattice is in both cases fcc. In full-Heusler alloy
the atomic basis consists of four atoms. For example, in
Co$_2$MnSi the positions of the basis atoms in Wyckoff coordinates
are the following: Co atoms at $(0\:0\:0)$ and
$({1\over2}\:{1\over2}\:{1\over2})$, Mn at
$({1\over4}\:{1\over4}\:{1\over4})$, Si at
$({3\over4}\:{3\over4}\:{3\over4})$. The Co atoms at the two
different sublattices have the same local environment rotated by
90$^o$ with respect to the [001] axis. In half-Heusler compounds
the position $({1\over2}\:{1\over2}\:{1\over2})$ is vacant.

The calculations are carried out with the augmented spherical
waves  method (ASW)\cite{asw} within the atomic-sphere
approximation (ASA).\cite{asa} The exchange-correlation potential
is chosen in the generalized gradient approximation. \cite{gga} A
dense Brillouin zone (BZ) sampling $30\times30\times30$ is used.
The radii of all atomic spheres are chosen equal. In the case of
half-Heusler alloys we introduce an empty sphere located at the
unoccupied site.

\subsection{Exchange parameters}

The method for the calculation of exchange constants has been
presented elsewhere.\cite{Sasioglou2004}
Here we give a brief overview.

We describe the interatomic exchange interactions in terms of
the classical Heisenberg Hamiltonian
\begin{equation}
\label{eq:hamiltonian2} H_{eff}=-  \sum_{\mu,\nu}\sum_{\begin
{array}{c}
^{{\bf R},{\bf R'}}\\ ^{(\mu{\bf R} \ne \nu{\bf R'})}\\
\end{array}} J_{{\bf R}{\bf R'}}^{\mu\nu}
{\bf e}_{\bf R}^{\mu}{\bf e}_{\bf R'}^{\nu}
\end{equation}
In Eq. (\ref{eq:hamiltonian2}), the  indices  $\mu$ and $\nu$
number different sublattices and ${\bf R}$ and ${\bf R'}$ are the
lattice vectors specifying the atoms within sublattices, ${\bf
e}_{\bf R}^\mu$ is the unit vector pointing in the direction of
the magnetic moment at site $(\mu,{\bf R})$.

We employ the frozen-magnon approach  to calculate interatomic
Heisenberg exchange parameters.\cite{magnon} The calculations
involve few steps. In the first step, the exchange parameters
between the atoms of a given sublattice $\mu$ are computed. The
calculation is based on the evaluation of the energy of the
frozen-magnon configurations defined by the following atomic polar
and azimuthal angles
\begin{equation}
\theta_{\bf R}^{\nu}=\theta, \:\: \phi_{\bf R}^{\nu}={\bf q \cdot
R}+\phi^{\nu}. \label{eq_magnon}
\end{equation}
The constant phase $\phi^{\nu}$ is  chosen equal to zero.
The magnetic moments of all other sublattices are kept parallel to
the z axis. Within the Heisenberg model~(\ref{eq:hamiltonian2})
the energy of such configuration takes the form
\begin{equation}
\label{eq:e_of_q} E^{\mu\mu}(\theta,{\bf
q})=E_0^{\mu\mu}(\theta)+\sin^{2}\theta J^{\mu\mu}({\bf q})
\end{equation}
where $E_0^{\mu\mu}$ does not depend on {\bf q} and the Fourier
transform $J^{\mu\nu}({\bf q})$ is defined by
\begin{equation}
\label{eq:J_q} J^{\mu\nu}({\bf q})=\sum_{\bf R} J_{0{\bf
R}}^{\mu\nu}\:\exp(i{\bf q\cdot R}).
\end{equation}

In the case of $\nu=\mu$ the sum in Eq. (\ref{eq:J_q}) does not
include ${\bf R}=0$. Calculating $ E^{\mu\mu}(\theta,{\bf q})$ for
a regular ${\bf q}$-mesh in the Brillouin zone of the crystal and
performing back Fourier transformation one gets exchange
parameters $J_{0{\bf R}}^{\mu\mu}$ for sublattice $\mu$. The
determination of the exchange interactions between the atoms of
two different sublattices $\mu$ and $\nu$  is discussed in Ref.
\onlinecite{Sasioglou2004}.

\subsection{Curie temperature}

The Curie  temperature is estimated within two different
approaches: the mean--field approximation (MFA) and random phase
approximation (RPA). The MFA for a multi-sublattice material
requires solving the system of  coupled
equations\cite{Sasioglou2004,Anderson}
\begin{equation}
\label{eq_system} \langle e^{\mu}\rangle
=\frac{2}{3k_BT}\sum_{\nu}J_0^{\mu\nu}\langle e^{\nu}\rangle
\end{equation}
where  $\langle e^{\nu}\rangle$ is the average $z$ component of
${\bf e}_{{\bf R}}^{\nu}$  and $J_0^{\mu\nu}\equiv\sum_{\bf R}
J_{0{\bf R}}^{\mu\nu}$. Eq. \ref{eq_system} can be represented in
the form of eigenvalue matrix-problem
\begin{equation}
\label{eq_eigenvalue} ({\bf \Theta}-T {\bf I}){\bf E}=0
\end{equation}
where $\Theta_{\mu\nu}=\frac{2}{3k_B}J_0^{\mu\nu}$, ${\bf I}$ is a
unit matrix and ${\bf E}$ is the vector of $\langle e^{\nu}\rangle
$. The largest eigenvalue of matrix $\Theta$ gives the value of
$T_C^{MFA}$.\cite{Anderson}

A more consequent method for the study of the thermodynamics of
Heisenberg systems is provided by the RPA approach.
\cite{tyablikov,Callen} The RPA technique is intensively used for
studies of both single-sublattice \cite{pajda,bouzerar}and
multi-sublattice\cite{antiferro,ferri,magnetite,Azaria,t_layer,m_layer_1,m_layer_2,Nolting,Turek2005}
systems. In the case that only the exchange interactions within
one sublattice are important the Curie temperature within the RPA
is given by the relation\cite{pajda}
\begin{equation}
\label{eq_RPA} \frac{1}{k_{B}T_{C}^{RPA}}=
\frac{3}{2}\frac{1}{N}\sum_q\frac{1}{J({\bf 0})- J({\bf q})},
\end{equation}
We use the RPA approach to study the temperature dependence of the
magnetization in the temperature interval from 0 K to $T_C$. The
RPA technique for a multi-sublattice system is briefly presented
in Appendix.

\begin{table}
\caption{Calculated atom-resolved and total spin moments in
$\mu_\mathrm{B}$ for NiMnSb, CoMnSb, Co$_2$CrAl and Co$_2$MnSi.
All compounds are half-metallic at the experimental lattice
constants taken from Ref. \onlinecite{Webster}.  a$_{II}$ means
the use of the lattice constant that places the Fermi level at the
upper edge of the half-metallic gap and a$_{III}$ corresponds to
1\% contraction of the lattice constant with respect to a$_{II}$.}
\label{table1}
\begin{ruledtabular}
 \begin{tabular}{lcccccc}
 Compound &  a(\AA ) &      X    &  Y     &   Z & Void& Total \\ \hline
 NiMnSb - a$_{I[exp]}$   &  5.93 &  0.20  &  3.85 & -0.09 & 0.04   &    4.00 \\
 NiMnSb - a$_{II}$       &  5.68 &  0.32  &  3.68 & -0.05 & 0.05   &    4.00 \\
 NiMnSb - a$_{III}$      &  5.62 &  0.33  &  3.64 & -0.04 & 0.05   &    3.97 \\ \hline
 CoMnSb - a$_{I[exp]}$   &  5.87 & -0.32  &  3.41 & -0.11 & 0.02   &    3.00 \\
 CoMnSb - a$_{II}$       &  5.22 &  0.45  &  2.57 & -0.06 & 0.04   &    3.00 \\
 CoMnSb - a$_{III}$      &  5.17 &  0.48  &  2.52 & -0.05 & 0.04   &    2.99 \\

\hline
Co$_2$CrAl - a$_{I[exp]}$ & 5.74 & 0.62  &   1.83 & -0.08 & -      &    3.00 \\
Co$_2$CrAl - a$_{II}$     & 5.55 & 0.69  &   1.68 & -0.06 & -      &    3.00 \\
Co$_2$CrAl - a$_{III}$    & 5.49 & 0.69  &   1.66 & -0.05 & -      &    2.99 \\
 \hline

Co$_2$MnSi - a$_{I[exp]}$ & 5.65 & 0.93  &   3.21 & -0.06 & -     &     5.00 \\
Co$_2$MnSi - a$_{II}$     & 5.49 & 0.97  &   3.10 & -0.04 & -     &     5.00 \\
Co$_2$MnSi - a$_{III}$    & 5.43 & 0.97  &   3.01 & -0.04 & -     &     4.97 \\

\end{tabular}
\end{ruledtabular}
\end{table}

\section{DOS and magnetic moments} \label{secIII}

\subsection{NiMnSb and CoMnSb}

In this section we report the calculation of DOS and magnetic
moments at different lattice  parameters for NiMnSb and for CoMnSb
compound that has one electron per formula unit less than NiMnSb.
The electronic structure of both compounds has been extensively
studied earlier and the reader is referred to the review
\onlinecite{GalanakisReview} and references therein for detailed
discussion. Here we present a brief description of the
calculational results aiming to provide the basis for further
considerations and to allow the comparison with previous work.

\begin{figure*}
  \begin{center}
    \includegraphics[scale=0.42]{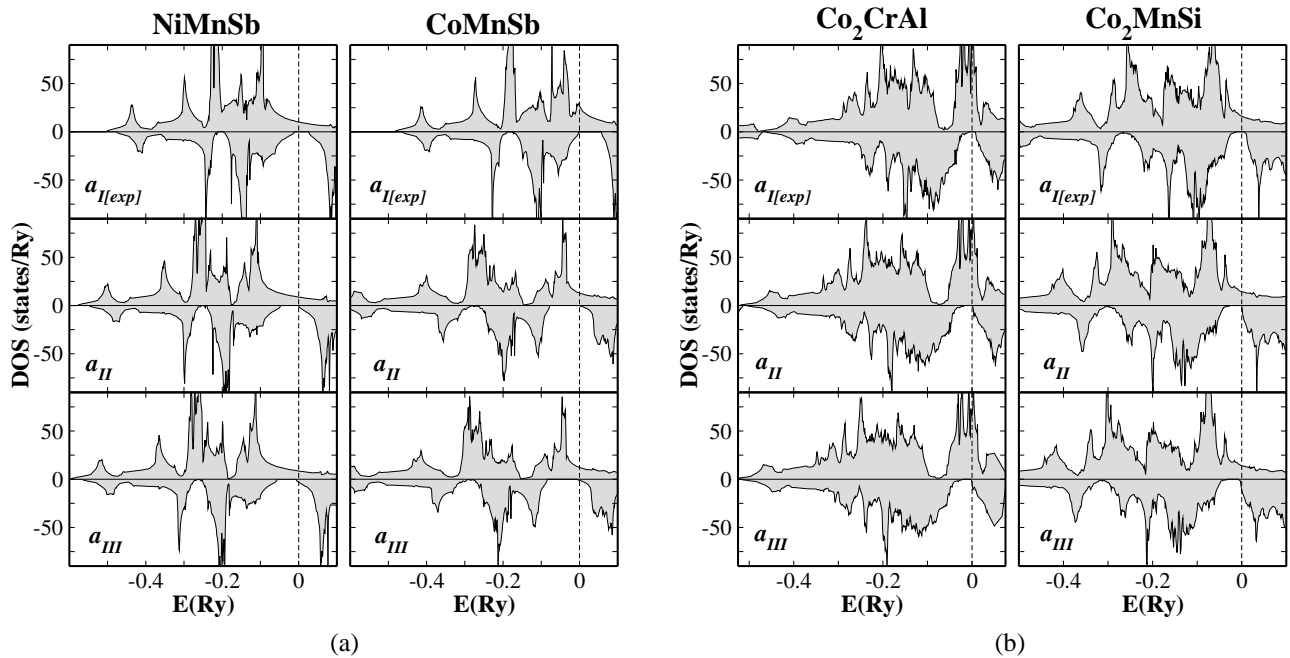}
  \end{center}
\caption{(a) Calculated spin-resolved density of states of NiMnSb
and CoMnSb for three values of the lattice parameter. (b) The same
for Co$_2$CrAl and  Co$_2$MnSi. The upper panels present the
results for the experimental lattice constant.\cite{Webster} The
middle panels show the results for the lattice parameter a$_{II}$
that is determined by the coincidence of the Fermi level with the
upper edge of the half-metallic gap. The bottom panels present the
results for lattice parameter a$_{III}$ that is obtained by a 1\%
contraction of a$_{II}$.} \label{fig2}
\end{figure*}

In Table \ref{table1} we collect the atomic and total spin moments
for three different lattice parameters. The investigation of the
influence of the value of the lattice parameter on the properties
of the Heusler alloys is important since the samples grown on
different substrates can have different lattice spacings. The
first calculation is performed for the experimental bulk lattice
constant.\cite{Webster} The calculated densities of states (DOS)
for this case are presented in the upper panel of Fig.~\ref{fig2}.
For both NiMnSb and CoMnSb the Fermi level lies in the low-energy
part of the half-metallic gap. The compression of the lattice
pushes the majority $p$ states to higher energies that results in
increased energy position of the Fermi level with respect to the
half-metallic gap. At the lattice parameter $a_{II}$ the Fermi
level coincides with the upper edge of the gap (Fig.~\ref{fig2}).
In the next step we further contracted the lattice constant by 1\%
(lattice parameter $a_{III}$, bottom panel in Fig.~\ref{fig2}). In
this case the Fermi level is slightly above the gap and the total
spin moment is slightly smaller than the integer values of 3 and 4
$\mu_\mathrm{B}$ for CoMnSb and NiMnSb respectively.

The contraction of the lattice leads to an increase of the
hybridization between the $d$ orbitals of different
transition-metal atoms. This results in a decrease of the spin
moment of Mn. In the case of NiMnSb this change is small: the
reduction of the Mn spin moment under lattice contraction from the
experimental lattice parameter to $a_{II}$ is $\sim$0.2 $\mu_B$.
The Ni spin moment increases by about the same value to preserve
the integer value of the total spin moment of 4 $\mu_\mathrm{B}$.

In CoMnSb, the half-metallic gap is larger than in NiMnSb. As a
result, the transition of the Fermi level to the upper gap-edge
requires a large lattice contraction of 11\% (Table~\ref{table1}).
This leads to a strong decrease of the Mn moment by
0.84$\mu_\mathrm{B}$. To compensate this decrease the Co moment
changes its sign transforming the magnetic structure from
ferrimagnetic to ferromagnetic.

The influence of the lattice contraction on the exchange
interactions and Curie temperature is discussed in the next
Section.

\subsection{Co$_2$CrAl and Co$_2$MnSi}

The second group of materials studied in the paper is formed by
the full-Heusler compounds Co$_2$MnSi and Co$_2$CrAl. The
electronic structure of these systems has been studied earlier.
\cite{GalanakisFull} Compared to half-Heusler systems, the
presence of two Co atoms per formula unit results in  an increased
coordination number of Co atoms surrounding Mn atoms (eight
instead of four in CoMnSb). This leads to  an increased
hybridization between the $3d$ orbitals of the Mn and Co atoms.
The spin moment of Co in Co$_2$MnSi is about 1 $\mu_\mathrm{B}$
that is considerably larger than the Co moment in CoMnSb. In
Co$_2$CrAl the Co moment is about 1/3rd smaller than in Co$_2$MnSi
that reflects a smaller value of the Cr moment compared to the Mn
moment (Table~\ref{table1}).

As in the case of the half-Heusler compounds discussed above, the
variation of the lattice parameter leads to the change in the
position of the Fermi level. At the experimental lattice parameter
the Fermi level of Co$_2$CrAl lies in the lower part of the
half-metallic gap while for Co$_2$MnSi it is close to the middle
of the gap (Fig. \ref{fig2}b). The contraction of the lattice
needed to place the Fermi level at the upper edge of the gap is
smaller than for CoMnSb. As a result, the change in the magnetic
moments is also relatively weak (Table \ref{table1}).

\section{Exchange  parameters and Curie temperature}
\label{secIV}

\subsection{NiMnSb and CoMnSb}

\begin{figure}[t]
\includegraphics[scale=0.4]{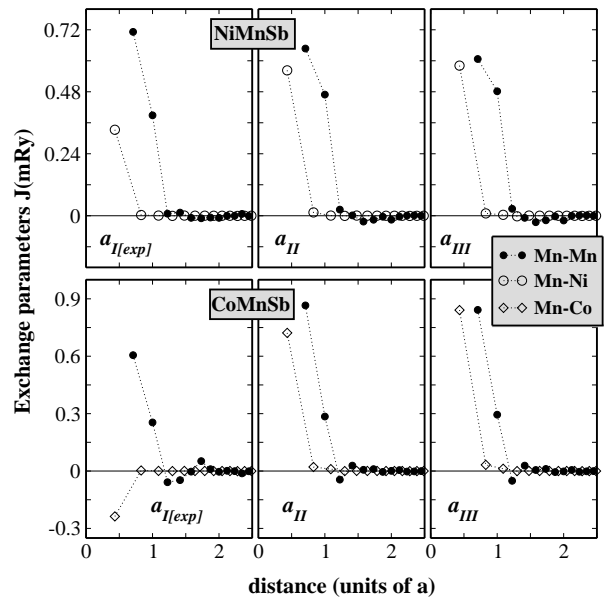}
\caption{The variation of the interatomic exchange parameters for
NiMnSb (upper panel) and CoMnSb (bottom panel) as a function of
the interatomic distance. The left panel corresponds to the
experimental lattice constant, the middle and right panels
correspond respectively to $a_{II}$ and $a_{III}$ parameters.}
\label{fig3}
\end{figure}

In Fig. \ref{fig3} we present the exchange constants calculated
for various lattice spacings. The Co-Co, Ni-Ni exchange
interactions as well as the exchange interactions between the
moments of the 3\textit{d} atoms and the induced moments of Sb
atoms are very weak and are not shown. The weakness of the
effective Co-Co and Ni-Ni exchange interactions can be explained
by a relatively large distance between atoms (Fig. \ref{fig1}) and
relatively small atomic moments.

On the other hand, each Ni(Co) atom is surrounded by four Mn atoms
as nearest neighbors that results in strong Mn-Ni(Co) exchange
interaction (Fig. \ref{fig3}). Also the exchange interaction
between large Mn moments is strong.

The ferromagnetic Mn-Mn interactions are mainly responsible for
the stable ferromagnetism of these materials. For both systems and
for all lattice spacings studied the leading Mn-Mn exchange
interaction is strongly positive. In NiMnSb, the Mn-Ni interaction
of the nearest neighbors is positive for all three lattice
parameters leading to the parallel orientation of the spins of the
Mn and Ni atoms. In CoMnSb the situation is different. At the
experimental lattice parameter the leading Mn-Co interaction is
negative resulting in the ferrimagnetism of the system. For the
contracted lattices the interaction changes sign resulting in the
ferromagnetic ground state of the alloy.

The analysis of the strength of the exchange interaction as a
function of the lattice parameter shows that in CoMnSb the
contraction leads to a strong increase of both leading Mn-Co and
Mn-Mn interactions. On the other hand, in NiMnSb the increase of
the Mn-Ni interaction is accompanied by a decrease of the leading
Mn-Mn interaction. Simultaneously, the interaction between the
second-nearest Mn atoms increases with contraction in the case of
NiMnSb staying almost unchanged in CoMnSb. This complexity of the
behavior reflects the complexity of the electronic structure of
the systems.

The interatomic exchange parameters are used to evaluate the Curie
temperature within two different approaches: MFA and RPA. In
Table~\ref{table2} we present the values of the Curie temperature
obtained, first, by taking into account the Mn-Mn interactions only
and, second, with account for both Mn-Mn and Mn-Ni(Co) interactions.
The  contribution of the inter-sublattice interactions to the
Curie temperature appears to be less than 5 \% for both compounds
and the Curie temperature is mainly determined by
the intra-sublattice Mn-Mn interaction.

The  MFA and RPA estimations of the Curie temperature differ
rather strongly (Table \ref{table2}). The relative difference of
two estimations is about 20\%. The reason behind this difference
will be discussed in the following section. For the systems
considered here the RPA estimations of the Curie temperatures are
in good agreement with the experiment, somewhat overestimating the
experimental values.

\begin{table}
\caption{Calculated Curie temperatures. The second and third
columns contain the $T_{C}^{MFA(RPA)}$ obtained with the account
for Mn-Mn (Cr-Cr) interactions only. In the next two columns all
interactions are taken into account. The last column presents the
experimental values of the Curie temperature from Ref.
\onlinecite{Webster}.} \label{table2}
\begin{ruledtabular}
 \begin{tabular}{lccccc}
$T_{C}$ (K) & MFA-Y & RPA-Y &
 MFA-all & RPA-all & Exp.
\\ \hline
NiMnSb - a$_{I[exp]}$     &  1096 & 880   &  1112 &  900 & 730  \\
NiMnSb - a$_{II}$         &  1060 & 853   &  1107 &  908 &  -   \\
NiMnSb - a$_{III}$        &  1008 & 802   &  1063 &  869 &  -   \\
\hline
CoMnSb - a$_{I[exp]}$     &  785  & 619   &  815  &  671  &  490 \\
CoMnSb - a$_{II}$         &  1185 & 940   &  1276 &  1052 &   -  \\
CoMnSb - a$_{III}$        &  1140 & 893   &  1252 &  1032 &   -  \\
\hline
Co$_2$CrAl - a$_{I[exp]}$ &   148 &  141  &  280  &  270  &  334 \\
Co$_2$CrAl - a$_{II}$     &   168 &  159  &  384  &  365  &   -  \\
Co$_2$CrAl - a$_{III}$    &   164 &  154  &  400  &  379  &   -  \\
\hline
Co$_2$MnSi - a$_{I[exp]}$ &   232 &  196  &   857  &  740  &  985 \\
Co$_2$MnSi - a$_{II}$     &   142 &  118  &   934  &  804  &   -  \\
Co$_2$MnSi - a$_{III}$    &   110 &  75   &   957  &  817  &   -  \\

\end{tabular}
\end{ruledtabular}
\end{table}

\begin{figure*}
\includegraphics[scale=0.45]{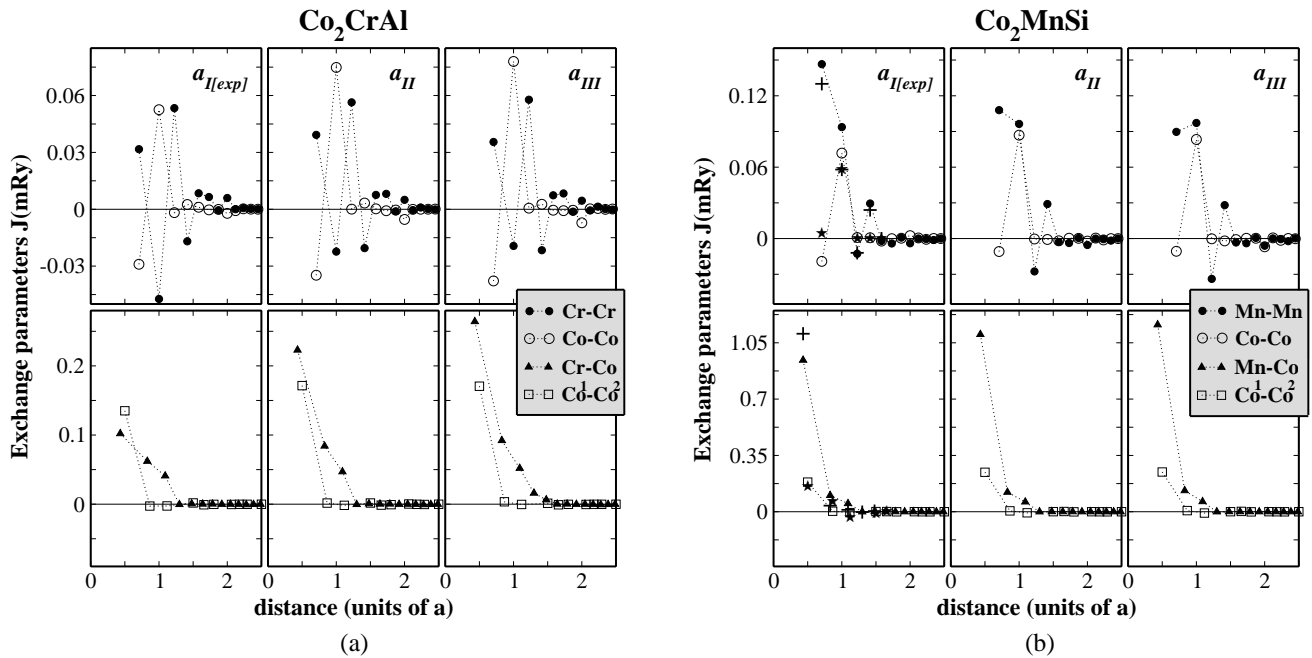}
\caption{(a) The exchange constants for Co$_2$CrAl  as a function
of the interatomic distance. (b) The same  for Co$_2$MnSi. The
left panels correspond to the experimental lattice constant, the
middle and right panels to  $a_{II}$ and  $a_{III}$ parameters
respectively. The superscripts 1 and 2 denote Co atoms belonging
to different sublattices (Fig. \ref{fig1}). For comparison, the
exchange parameters of Co$_2$MnSi obtained in Ref.
\onlinecite{Kurtulus} at the experimental lattice parameter
($a_{I[exp]}$) are shown. The following symbols are used in the
presentation: + for the Mn-Mn and Mn-Co interactions and $\star$
for the Co-Co and Co$^1$-Co$^2$ interactions. } \label{fig4}
\end{figure*}

Recently  K\"ubler \cite{Kubler2003} reported estimations of the
Curie temperature of NiMnSb. His approach is based on the
evaluation of the non-uniform magnetic susceptibility on the basis
of the Landau-type expansion for the free energy. Within some
approximations the parameters used in the study of the
thermodynamical properties can be expressed in terms of the
quantities evaluated within the first-principles DFT calculations.
The estimated values of the Curie temperature are 601 K for a
static approach and 701 K if the frequency dependence of the
susceptibility is taken into account. These estimations are
somewhat lower than the value of 880 K given by the RPA approach
(Table~\ref{table2}). A detailed comparative analysis of the two
calculational schemes is needed to get an insight in the physical
origin of this difference.

The contraction of the lattice in the case of the NiMnSb compound
leads to an increase of the Mn-Ni interactions (Fig.~\ref{fig3}).
This results in increased difference between the Curie
temperatures calculated with the Mn-Mn interactions only and with
both Mn-Mn and Ni-Mn interactions taken into account
(Table~\ref{table2}). For CoMnSb, the leading exchange
interactions of both Mn-Mn and Mn-Co types increase in the value
under transition from the experimental lattice constant to
$a_{II}$ (Fig.~\ref{fig3}). As a result, the Curie temperature
increases with contraction by about 50\%.

\subsection{Co$_2$CrAl and Co$_2$MnSi}

The presence of an extra Co atom in the full-Heusler alloys makes
the interactions more complex than in the case of the half-Heusler
alloys. In CoMnSb the important interactions arise between nearest
Mn atoms (Mn-Mn interactions) and between nearest Mn and Co atoms
(Mn-Co interaction). In the case of Co$_2$MnSi (Fig. \ref{fig4})
the interactions between Co atoms at the same sublattice (Co-Co)
and between Co atoms at different sublattices (Co$^1$-Co$^2$) must
be taken into account. The cobalt atoms at different sublattices
have the same local environment rotated by 90$^o$ about the [001]
axis. The leading interaction responsible for the stability of the
ferromagnetism is the Mn-Co interaction between Mn atoms and eight
nearest Co atoms (Fig. \ref{fig4}). This interaction changes
weakly with the contraction of the lattice. Our exchange
parameters agree well with the parameters of Kurtulus \textit{et al.}
(Fig.~\ref{fig4}) who also found the Co-Mn exchange interaction to
be leading. \cite{Kurtulus}

The interaction between nearest Co atoms at different sublattices
(empty squares in Fig.~\ref{fig4}) favors the ferromagnetism also
and is stronger than the ferromagnetic interaction between the
nearest Mn atoms (filled spheres). Although the spin moment of Mn
atoms is larger than the moment of Co atoms (Table~\ref{table1})
the opposite relation between exchange parameters can be the
consequence of the smaller distance between the Co atoms:  $a/2$
between the Co atoms and $\sqrt{2}a/2$ between the Mn atoms. An
interesting feature of the intra-sublattice Mn-Mn and Co-Co
interactions is different signs of the exchange parameters for
different distances between atoms. This leads to a RKKY-like
oscillations of the parameters (Fig.~\ref{fig4}).

In Co$_2$CrAl the leading Cr-Co interactions (filled triangles)
are much smaller than corresponding Mn-Co interactions in
Co$_2$MnSi. On the other hand, the leading inter-sublattice
ferromagnetic Co-Co interactions are comparable in both systems.
The compression of the lattice leads to an increase of the
magnitude of the inter-sublattice Co-Cr and Co$^1$-Co$^2$
coupling. The intra-sublattice Cr-Cr and Co-Co interactions
oscillate with varying inter-atomic distances.

The difference in the properties of the exchange parameters of the
half- and full-Heusler alloys is reflected in the calculated Curie
temperatures (Table~\ref{table2}). In contrast to CoMnSb where the
Mn-Mn exchange  interactions are dominant, in Co$_2$MnSi they play
a secondary role. The T$_{C}^{MFA(RPA)}$ calculated taking into
account these interactions only is much smaller than the Curie
temperature calculated with all inter-atomic exchange interactions
taken into account (Table \ref{table2}). The same conclusion is
valid for Co$_2$CrAl where the Cr-Cr interactions give about half
of the Curie temperature obtained with all interactions included
into consideration.

\begin{figure}
\includegraphics[scale=0.42]{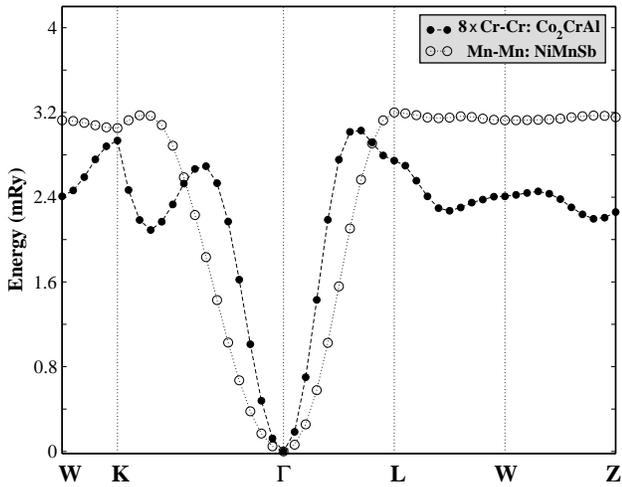}
\caption{The frozen-magnon dispersions for NiMnSb and Co$_2$CrAl
along the symmetry lines in the Brillouin zone. The magnons
correspond to the Mn sublattice in the case of NiMnSb and to the
Cr sublattice in the case of Co$_2$CrAl. For Co$_2$CrAl, the
frozen-magnon energies are multiplied by a factor of 8.}
\label{fig5}
\end{figure}

\begin{figure}
\includegraphics[scale=0.44]{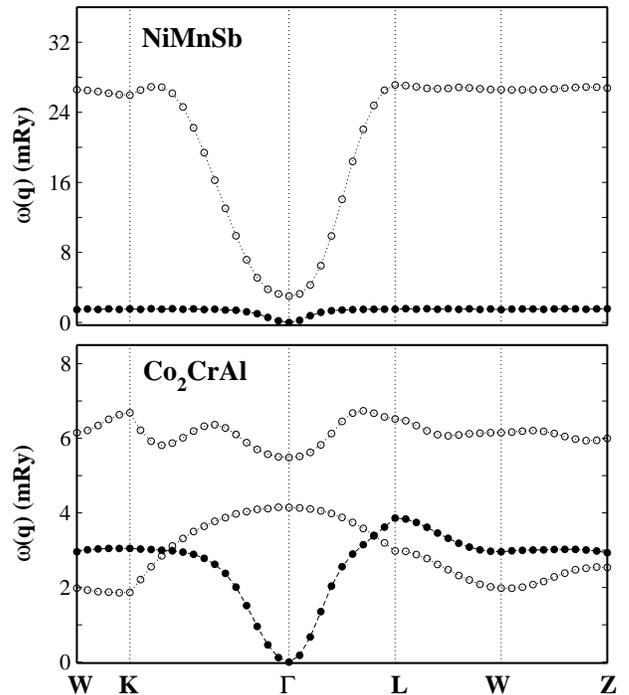}
\caption{The spin wave dispersions for NiMnSb and Co$_2$CrAl along
the symmetry lines in the Brillouin zone. Filled and empty spheres
denote the  acoustic  and  optical branches, respectively.}
\label{fig6}
\end{figure}

A striking feature of the full Heusler compound  Co$_2$CrAl that
differs it strongly from the half-Heusler systems considered in
the previous Section is a very small difference between the $T_C$
values calculated within the MFA and RPA approaches. A similar
behavior was obtained  for the Curie temperatures of the
zincblende  MnSi and MnC.\cite{stability_zb} In Co$_2$MnSi, the
relative difference of the MFA and RPA estimations assumes an
intermediate position between the half-Heusler systems and
Co$_2$CrAl.

To understand the origin of the strong variation of the relative
difference of the MFA and RPA estimations of the Curie temperature
we compare in Fig.~\ref{fig5} the frozen magnon dispersions for
two compounds. The magnons correspond to the Mn sublattice in the
case of NiMnSb and to the Cr sublattice in the case of Co$_2$CrAl.
As seen from Table~\ref{table2} the MFA and RPA estimations
obtained with the use of these dispersions differ by 20\% for
NiMnSb and by 5\% for Co$_2$CrAl.

The Curie temperature is given by the average value of the magnon
energies. In MFA this is the arithmetic average while in RPA this
is harmonic average. Therefore we need to understand why for
Co$_2$CrAl these two averages are much closer than for NiMnSb. The
following properties of the averages are important for us.  The
arithmetic average takes all the magnon values with equal weight
whereas in the harmonic average the weight decreases with
increasing energy of the
magnon.\cite{pajda,bouzerar,stability_zb,above_room,sabiryanov} It
is an arithmetic property that the MFA estimation is larger than
the RPA one or equal to it if all numbers to be averaged are equal
to each other. In terms of magnon energies, $T_{C}^{MFA}$ is equal
to $T_{C}^{RPA}$ in the case that the magnon spectrum is
dispersion-less.

Considering the frozen-magnon dispersions from the viewpoint of
these properties we indeed can expect that the arithmetic and
harmonic averages will be closer for Co$_2$CrAl. In
Fig.~\ref{fig5} both curves are scaled to have almost the same
maximal value. It is seen that the Co$_2$CrAl dispersion has
smaller relative contribution of the low-energy magnons because of
the steeper increase of the curve at small wave vectors. It has
also smaller contribution of the magnons with the largest energies
because the maxima have the form of well-defined peaks opposite to
NiMnSb where we get a plateau. Thus the main contribution in the
case of  Co$_2$CrAl comes from intermediate energies that makes
the MFA and RPA estimations closer.

In Fig.~\ref{fig6} we present the calculated spin-wave spectra for
NiMnSb and Co$_2$CrAl. The spin-wave energies are obtained by the
diagonalization of the matrix of exchange parameters that contains
all important intra- and inter-sublattice interactions. The number
of branches in the spectrum is equal to the number of magnetic
atoms in the unit cell: two in NiMnSb and three in Co$_2$CrAl. One
of the branches is acoustic and has zero energy for zero wave
vector. Also in the spin-wave spectra, we see strong difference
between two systems. In  NiMnSb, the acoustic branch is
predominantly of the Ni type stemming from the weak interaction
between Mn and Ni magnetic moments (see Fig.~\ref{fig3}). On the
other hand, the optical branch is of predominantly the Mn type.
The strong hybridization between two sublattices is obtained only
about $\textbf{q}=0$. In Co$_2$CrAl, the energy scale of the
branches differs much smaller and the hybridization between
sublattices is stronger than for  NiMnSb.

Coming back to the considerations of the Curie temperatures, we
conclude that, in general, the Curie temperatures of  Co$_2$MnSi
and Co$_2$CrAl calculated within both MFA and RPA are in good
agreement with experiment while the MFA values in the case of
NiMnSb and CoMnSb overestimate the Curie temperature strongly.

The lattice contraction leads in both compounds to an enhancement
of the Mn-Co(Cr-Co) exchange constants that results in an increase
of the Curie temperature.

Kurtulus and collaborators have calculated the Curie temperature
for Co$_2$MnSi within MFA and found the value of 1251 K that is
considerably larger than our MFA estimate of 857 K. This
difference is unexpected since the values of the exchange
parameters obtained by Kurtulus \textit{et al.} agree well with
our parameters (Fig.\ref{fig4}). To reveal the origin of the
discrepancy we performed the MFA calculation of the Curie
temperature with the exchange parameters of Kurtulus \textit{et
al.} and obtained the $T_C$ value of 942 K which is in reasonable
agreement with our estimate. Apparently the reason for the
inconsistency is in the procedure of the solving of the
multiple-sublattice MFA problem used by Kurtulus \textit{et al.}
that should deviate from the standard one. \cite{Anderson}

\section{Temperature dependence of the  magnetization}
\label{secV}

\begin{figure*}
\begin{flushleft}
\includegraphics[scale=0.44]{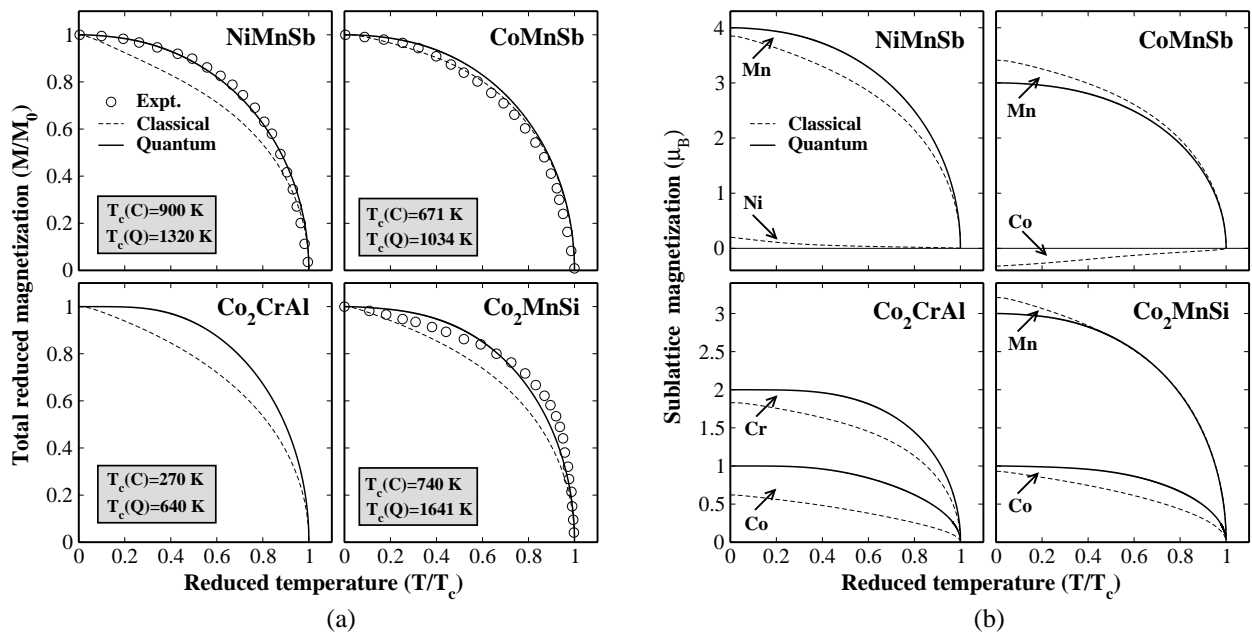}
\caption{(a) The calculated temperature dependence of the total
magnetization for both families of Heusler  alloys. For comparison
the experimental temperature dependences\cite{Webster} are
presented. The calculations are performed for both classical and
quantum Hamiltonians. Both the magnetization and the temperature
are given in reduced form. (b) Calculated sublattice
magnetizations as a function of temperature. The temperature is
given in reduced form.} \label{fig7}
\end{flushleft}
\end{figure*}

The study of the temperature dependence of the magnetic properties
of itinerant ferromagnets is one of the fundamental problems of
ongoing researches. Although density functional theory can
formally be extended to the finite temperatures \cite{Mermin}, it
is rarely used because of the lack of suitable
exchange-correlation potentials  for magnetic systems at finite
temperatures. Statistical mechanics treatment of model
Hamiltonians is usually employed. In this section we will present
the results of the calculation of the temperature dependence of
magnetization that is based on the consideration of the Heisenberg
hamiltonian with exchange parameters calculated within a
parameter-free DFT approach (Sect.~\ref{secII}).

To calculate the temperature dependence of the  magnetization we
use the RPA  method as  described  in appendix. We consider both
classical-spin and quantum-spin cases.

In the classical-spin calculations the calculated values of the
magnetic moments (Table~\ref{table1}) are used. To perform
quantum-mechanical RPA calculation we assign integer values to the
atomic moments. In the semi Heusler compounds we ignore the
induced moments on Ni and Co atoms and assign the whole moment per
formula unit to the Mn atom: 4$\mu_B$ ($S=2$) in NiMnSb and
3$\mu_B$ ($S=3/2$) in CoMnSb. In Co$_2$MnSi we take the values of
3$\mu_B$ ($S=3/2$) and 1$\mu_B$ ($S=1/2$) for Mn and Co atoms
respectively. This assignment preserves the value of the total
spin moment per chemical unit. In Co$_2$CrAl we use in the
quantum-RPA calculations the atomic moment of 2$\mu_B$ ($S=1$) for
Cr and 1$\mu_B$ ($S=1/2$) for Co.

In Fig.~\ref{fig7}(a), we  present in the normalized form the
calculated temperature dependence  of the magnetization for both
families of Heusler compounds. The calculations are performed for
the experimental lattice parameter. For comparison, the
experimental curves are presented. The nature of the spin (quantum
or classical) influences the form of the curves considerably. The
classical curve lies lower than the quantum one. This results from
a faster drop of the magnetization in the low-temperature region
in the case of classical spins. In general, the quantum
consideration gives better agreement of the form of the
temperature dependence of the magnetization with experiment.

In Fig.~\ref{fig7}(b) we present the temperature dependence of the
magnetization of individual sublattices. As expected from the
previous discussions in half-Heusler systems the main contribution
to the magnetization comes from the Mn sublattice while for the
full-Heusler systems both 3\textit{d} atoms contribute
substantially.

\begin{figure}
\includegraphics[scale=0.42]{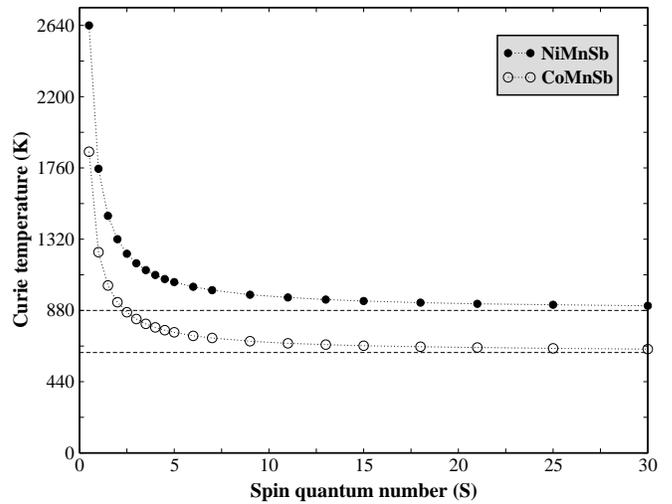}
\caption{Curie temperature of NiMnSb and CoMnSb as a function of
spin quantum number $S$. The horizontal broken lines correspond to
the classical limit ($S =\infty$).} \label{fig8}
\end{figure}

Considering the calculated Curie temperatures we notice that the
value of $T_C$ calculated within the quantum-mechanical RPA is
substantially larger than the corresponding classical estimation
(see Fig.~\ref{fig7}). This property is well-known and has its
mathematical origin in the factor $(S+1)/S$ entering the RPA
expression for the Curie temperature (Eq.~\ref{langevin_quantum}).
In Fig.~\ref{fig8} we show the dependence of the Curie temperature
calculated within the quantum mechanical RPA approach on the value
of $S$. The exchange parameters are kept unchanged in these
calculations. We see that the dependence has a monotonous
character tending to a classical limit for large $S$.

Presently we do not have an explanation why quantum-mechanical
calculations give better form of the temperature dependence while
the classical calculation provides better value of the Curie
temperature. We can suggest the following arguments. The quantum
treatment is more appropriate than the classical one in the
low-temperature region. At high temperatures characterized by
strong deviation of the atomic spins from the magnetization axis
the quantum treatment gives too slow decrease of the
magnetization. It is worth noting that the consequent theory
should take into account not only the orientational disorder of
the atomic moments but also the single-particle (Stoner-type)
excitations leading to the decrease of atomic moments. Another
important aspect is related to the fact that the exchange
parameters used in the calculations are estimated within the
picture of classical atomic moments described above. It is
possible that the values of the exchange parameters must be
modified for the use in the quantum-mechanical case. These
questions belong to fundamental problems of the quantum-mechanical
description of the magnetic systems with itinerant electrons.

\section{Summary and conclusions}\label{secVI}

We studied the electronic structure of several Heusler alloys
using the augmented spherical waves method in conjunction with the
generalized gradient approximation to the exchange and correlation
potential. Using the frozen-magnon approximation we calculated
inter-atomic exchange parameters that were used to estimate the
Curie temperature. The Curie temperature was estimated within both
mean-field and random-phase approximation techniques.

For the half-Heusler alloys NiMnSb and CoMnSb the dominant
interaction is between the Mn atoms. The lattice compression
results in considerable change of the exchange parameters and
Curie temperature.

The magnetic interactions are more complex in full-Heusler alloys
Co$_2$MnSi and Co$_2$CrAl. In both cases the ferromagnetism is
stabilized  by the inter-sublattice interactions between the
Mn(Cr) and Co atoms and between Co atoms belonging to different
sublattices. Both the random phase and mean field approximations
slightly underestimate the values of the Curie temperature.
Compression of the lattice constant has little effect on the
magnetic properties of the full-Heusler alloys.

We study the temperature dependence of the magnetization within
the quantum mechanical and classical RPA. The quantum-mechanical
approach gives the form of the temperature dependence that is in
good agreement with experiment. The value of the Curie temperature
is, however, overestimated in the quantum-mechanical calculation.

\begin{acknowledgments}
The financial support of Bundesministerium f\"ur
Bildung und Forschung is acknowledged. I.G. is a fellow of the
Greek State Scholarship Foundation. We thank the authors of
Ref. \onlinecite{Turek2005} for making the manuscript available
before publication.
\end{acknowledgments}

\appendix
\section{The random phase approximation for multi-sublattice
Heisenberg Hamiltonian}

The Green function approach is a powerful tool in the study of the
magnetism of complex systems. (See, e.g., the application of the
method to antiferromagnet\cite{antiferro},
ferrimagnets\cite{ferri,magnetite}, random alloys\cite{Azaria},
layered systems \cite{t_layer,m_layer_1,m_layer_2}, disordered
dilute magnetic systems \cite{Nolting}, multi-sublattice
ferromagnets \cite{Turek2005}.) In this appendix we briefly
overview the formalism to study the temperature dependence of the
magnetization of multi-sublattice systems within the random phase
approximation.

We start with the Heisenberg Hamiltonian for quantum spins
\begin{equation}
\label{eq:hamiltonian3} H=-\sum_{ij} \sum_{\mu\nu} J_{ij}^{\mu\nu}
{\bf e}_{i,\mu}{\bf e}_{j,\nu}
\end{equation}
where ${\bf
e}_{i,\mu}=(\hat{s}_{i,\mu}^{x},\hat{s}_{i,\mu}^{y},\hat{s}_{i,\mu}^{z})/(S_{\mu})$
is the  normalized  spin operator  corresponding to site
($i,\mu$).

In terms of the  creation and destruction  operators $
\hat{s}_{i,\mu}^{\mp}=\hat{s}_{i,\mu}^{x}\mp \hat{s}_{i,\mu}^{y}$
the Hamiltonian can be  written in the   form
\begin{equation}
\label{eq:hamiltonian4} H=-\sum_{ij}\sum_{\mu\nu}
\tilde{J}_{ij}^{\mu\nu} [ \hat{s}_{i,\mu}^{+}\hat{s}_{j,\nu}^{-} +
\hat{s}_{i,\mu}^{z} \hat{s}_{j,\nu}^{z}]
\end{equation}
where $\tilde{J}_{ij}^{\mu\nu}=J_{ij}^{\mu\nu}/{S_{\mu}S_{\nu}}$.

Following Callen \cite{Callen} let us introduce Green function
\begin{equation}
G_{ij}^{\mu\nu}(\tau)=-\frac{i}{\hbar}\theta(\tau)\langle[\hat{s}_{i,\mu}^{+}(\tau),
\exp(\eta \hat{s}_{j,\nu}^{z})\hat{s}_{j,\nu}^{-}]\rangle
\label{mixed}
\end{equation}
where   $\eta$  is a  parameter, $\theta(\tau)$ is the   step
function ($\theta(\tau)=1$ for $\tau\geq 0$),     $[\ldots]$
denotes the commutator and $\langle\ldots\rangle$ is the  thermal
average over the canonical ensemble, ie., $\langle F
\rangle=\textrm{Tr}[ \exp(-\beta H)F]/\textrm{Tr}[ \exp(-\beta
H)]$ with $\beta=1/k_{B}T$

Writing the  equation of motion  for $G_{ij}^{\mu\nu}(\tau)$ we
obtain
\begin{eqnarray}
\frac{\partial}{\partial \tau}G_{ij}^{\mu\nu}(\tau)& = &
-\frac{i}{h}\delta(\tau) \langle[\hat{s}_{i,\mu}^{+}(\tau),
\exp(\eta \hat{s}_{j,\nu}^{z})\hat{s}_{j,\nu}^{-}]\rangle - \frac{1}{\hbar^{2}}\theta(\tau) \nonumber  \\
& & \times \langle[[\hat{s}_{i,\mu}^{+}(\tau), \hat{H}], \exp(\eta
\hat{s}_{j,\nu}^{z})\hat{s}_{j,\nu}^{-}]\rangle \label{e_motion}
\end{eqnarray}
The last  commutator  term  in Eq.~(\ref{e_motion}) generates
higher-order Green functions. These functions can be  reduced to
lower-order functions by using  Tyablikov decoupling (random phase
approximation) scheme\cite{tyablikov}:
\begin{equation}\label{tyablikov}
\langle[\hat{s}_{i,\mu}^{+}(\tau) \hat{s}_{k,\mu}^{z}  ,
\hat{s}_{j,\nu}^{-}]\rangle \approx \langle\hat{s}_{k,\mu}^{z}
\rangle \langle[\hat{s}_{i,\mu}^{+}(\tau),
\hat{s}_{j,\nu}^{-}]\rangle
\end{equation}
Applying this  decoupling procedure to Eq.~(\ref{e_motion}) we get
\begin{eqnarray}
\frac{\partial}{\partial \tau}G_{ij}^{\mu\nu}(\tau)& = &
-\frac{i}{h}\delta(\tau) \langle[\hat{s}_{i,\mu}^{+}(\tau),
\exp(\eta \hat{s}_{j,\nu}^{z})\hat{s}_{j,\nu}^{-}]\rangle \nonumber\\
&&+ \frac{2i}{\hbar} \sum_{k,\xi}\tilde{J}_{i,k}^{\mu\xi}[\langle
\hat{s}_{i,\mu}^{z}\rangle G_{kj}^{\xi\nu}(\tau) \nonumber\\
&&- \langle \hat{s}_{k,\xi}^{z}\rangle G_{ij}^{\mu\nu}(\tau)]
\label{e_motion3a}
\end{eqnarray}
After a  Fourier transformation to energy  and  momentum space
[$g({\bf q},\omega)=\frac{1}{2\pi}\sum_{l}\int d \omega e^{-i{\bf
q}{ \bf R}_{l}}G_{l0}(\tau)$] we  obtain
\begin{eqnarray}
\hbar \omega g_{\mu\nu}( {\bf q},\omega)& = & \frac{1}{2\pi}
\langle[\hat{s}_{\mu}^{+},\exp(\eta \hat{s}_{\nu}^{z})\hat{s}_{\nu}^{-}]\rangle \delta_{\mu\nu} \nonumber\\
& & -2\sum_{\xi} \{ \tilde{J}_{\mu\xi}({\bf q})\langle
\hat{s}_{i,\mu}^{z}\rangle g_{\xi\nu}({\bf q},\omega) \nonumber\\
& & -\tilde{J}_{\mu\xi}({\bf 0})\langle \hat{s}_{k,\xi}^{z}\rangle
g_{\mu\nu}({\bf q},\omega)\} \label{e_motion3}
\end{eqnarray}
Eq.~(\ref{e_motion3})  can  be  written in a compact matrix form
\begin{equation}
[\hbar\omega \textbf{I}- \textbf{M}({\bf q})]\textbf{g}({\bf
q},\omega)=\textbf{u}
\label{matrix}
\end{equation}
where $ \textbf{g}({\bf q},\omega)$ is  a  symmetric square
matrix, \textbf{I} is a unit matrix and the inhomogeneity matrix
${\bf u}$ is expressed by
\begin{equation}
u_{\mu\nu}=\frac{1}{2\pi}\langle[\hat{s}_{\mu}^{+},\exp(\eta
\hat{s}_{\nu}^{z})\hat{s}_{\nu}^{-}]\rangle\delta_{\mu\nu},
\end{equation}
matrix $\textbf{M}({\bf q}) $ is defined by
\begin{equation}
M_{\mu\nu}({\bf q})=\bigg \{ \sum_{\xi}2\tilde{J}_{\mu\xi}({\bf
0})\langle\hat{s}_{\xi}^{z} \rangle \bigg\}\delta_{\mu\nu}-
2\tilde{J}_{\mu\nu}({\bf q})\langle\hat{s}_{\mu}^{z} \rangle
 \label{m_elements}
\end{equation}

Next, we introduce the transformation which diagonalizes matrix
${\bf M}({\bf q})$:\cite{m_layer_2}
\begin{equation}\label{dioganal}
{\bf L}({\bf q}){\bf M}({\bf q}){\bf R}({\bf q})=\Omega({\bf q})
\end{equation}
where $\Omega({\bf q})$ is  the diagonal matrix whose elements
give the spin wave energies $\omega_{\mu}({\bf q})$. The number of
branches in the spin wave spectrum is equal to the number of
magnetic atoms in the unit cell. The transformation matrix ${\bf
R}({\bf q})$ and  its inverse ${\bf R}^{-1}({\bf q})={\bf L}({\bf
q})$ are obtained from the right eigenvectors of ${\bf M}({\bf
q})$ as columns and from the left eigenvectors as rows,
respectively.

Using  the spectral theorem  and  Callen's  technique
\cite{Callen} one  obtains the thermal averages of the
sublattice magnetizations:
\begin{equation}\label{callen}
\langle
\hat{s}_{\mu}^{z}\rangle=\frac{(S_{\mu}-\Phi_{\mu})(1+\Phi_{\mu})^{2S_{\mu}+1}+(S_{\mu}+1+\Phi_{\mu})\Phi_{\mu}
^{2S_{\mu}+1}}{(1+\Phi_{\mu})^{2S_{\mu}+1}-(\Phi_{\mu})^{2S_{\mu}+1}}
\end{equation}
where $\Phi_{\mu}$ is an auxiliary function given by
\begin{equation}\label{phi_function}
\Phi_{\mu}=\frac{1}{N}\sum_{q}\sum_{\nu}L_{\mu\nu}({\bf
q})\frac{1}{e^{\beta \omega_{\nu}({\bf q})}-1}R_{\mu\nu}({\bf q})
\end{equation}
In Eq.~(\ref{phi_function}), $N$ is the number of ${\bf q}$ points
in the  first BZ.

Eq.~(\ref{callen}) is  the central equation for the calculation of
the sublattice magnetizations. It must be  solved
self-consistently. The Curie temperature $T_C$ is determined as
the point  where all sublattice magnetizations vanish.

Near $T_C$ ($\Phi_{\mu} \rightarrow \infty$ and $\langle
\hat{s}_{\mu}^{z} \rangle \rightarrow 0 $) Eq.~(\ref{callen}) can
be simplified. Expanding in $\Phi_{\mu}$ and using
Eq.~(\ref{phi_function}) one obtains
\begin{equation}\label{langevin_quantum}
\langle \hat{s}_{\mu}^{z}\rangle=\frac{(S_{\mu}+1)}{3S_{\mu}}
\bigg\{ \frac{1}{S_{\mu}^{2}N}\sum_{q,\nu}L_{\mu\nu}({\bf
q})\frac{1}{e^{\beta \omega_{\nu}({\bf q})}-1}R_{\mu\nu}({\bf
q})\bigg\}^{-1}.
\end{equation}
From Eq.~(\ref{langevin_quantum}), it follows that for
spin-independent Heisenberg exchange parameters
[Eq.~(\ref{eq:hamiltonian3})] the dependence of the Curie
temperature on the spin value is defined by the factor
$(S_{\mu}+1)/S_{\mu}$.

The classical limit  can be obtained by letting
$S_{\mu}\rightarrow \infty$ in Eqs.~(\ref{callen}) and
(\ref{langevin_quantum}).\cite{Turek2005}  Factor
$(S_{\mu}+1)/S_{\mu}$ in Eq.~(\ref{langevin_quantum}) becomes in
this case unity. The temperature dependence of the magnetization
can be calculated using a semiclassical analog of
Eq.~(\ref{callen}) given by \cite{Turek2005,Callen2}
\begin{equation}\label{langevin}
\langle {e}_{\mu}^{z}\rangle= \mathcal{L} \bigg(  \bigg\{
\frac{1}{N}\sum_{q,\nu}L_{\mu\nu}({\bf q})\frac{1}{e^{\beta
\omega_{\nu}({\bf q})}-1}R_{\mu\nu}({\bf q})\bigg\}^{-1}\bigg)
\end{equation}
where $\mathcal{L}(x)=\coth(x)-1/x$ is the  Langevin function and
$ {\bf e}_{\mu}$ is  the angular momentum vector of size one.

\end{document}